\documentclass[useAMS,usenatbib]{mn2e}

\usepackage{graphicx}

\def\iphas{IPHASXJ194359.5+170901}
\def\deg{$^\circ$}
\def\kms{\relax \ifmmode {\,\rm km\,s}^{-1}\else \,km\,s$^{-1}$\fi}
\def\td{$tD^{-1}$}
\def\ne{$N_e$}
\def\te{$T_e$}

\title[\iphas]{The Necklace: equatorial and polar outflows from the binary
  central star of the new planetary nebula \iphas}
\author[R.L.M. Corradi et al.]{R.L.M. Corradi$^{1,2}$\thanks{E-mail:
    rcorradi@iac.es}, L. Sabin$^{3}$, B. Miszalski$^{4}$,
  P. Rodr{\'{\i}}guez--Gil$^{5,1,2}$, \newauthor
  M. Santander--Garc{\'{\i}}a$^{5,1,2}$, D. Jones$^{6}$, J. Drew$^{4}$, 
  A. Mampaso$^{1,2}$, M. Barlow$^{7}$, \newauthor M.M. Rubio--D\'\i
  ez$^{5,8}$, J. Casares$^{1,2}$, K. Viironen$^{9,10}$, D.J. Frew$^{11}$, 
  \newauthor   C. Giammanco$^{1,2}$, R. Greimel$^{12}$, S. Sale$^{13,14}$\\
$^{1}$Instituto de Astrof{\'{\i}}sica de Canarias, E-38200 La Laguna,
Tenerife, Spain\\
$^{2}$Departamento de Astrof{\'{\i}}sica, Universidad de La Laguna,
E-38206 La Laguna, Tenerife, Spain\\
$^{3}$Instituto de Astronom\'{i}a, Universidad Nacional Aut\'{o}noma de 
M\'{e}xico, Apdo. Postal 877, 22800 Ensenada, B.C, Mexico\\
$^{4}$Centre for Astrophysics Research, STRI, University of
Hertfordshire, College Lane Campus, Hatfield AL10 9AB, UK\\
$^{5}$Isaac Newton Group of Telescopes, Apart. de Correos 321, 38700
Santa Cruz de la Palma, Spain\\
$^{6}$Jodrell Bank Centre for Astrophysics, School of Physics and Astronomy,
University of Manchester, M13 9PL, UK\\ 
$^{7}$Department of Physics and Astronomy, University College London, 
Gower Street, London WC1E 6BT, UK\\ 
$^{8}$Centro de Astrobiolog\'\i a, CSIC-INTA, Ctra de Torrej\'on a Ajalvir km 4,
E-28850 Torrej\'on de Ardoz, Spain\\ 
$^{9}$Centro Astron\'omico Hispano Alem\'an, Calar Alto, C/Jes\'us Durb\'an 
Rem\'on 2-2, E-04004 Almeria, Spain\\
$^{10}$Centro de Estudios de F\'\i sica del Cosmos de Arag\'on (CEFCA), 
C/General Pizarro 1-1, E-44001 Teruel, Spain\\
$^{11}$Department of Physics and Astronomy, Macquarie University, North Ryde,  
NSW 2109,  Australia\\
$^{12}$Institut f\"ur Physik, Karl-Franzen Universit\"at Graz, 
Universit\"atsplatz 5, 8010 Graz, Austria\\
$^{13}$Dept. de F\'\i sica y Astronom\'\i a,  
Universidad de Valpara\'\i so, Ave. Gran Bretaña 1111, Playa Ancha, 
Casilla 53,Valpara\'\i so, Chile\\
$^{14}$Dept. de Astronom\'\i a y Astrof\'\i sica, Pontificia Universidad
Cat\'olica de Chile, Av. Vicu\~na Mackenna 4860, Casilla 306, Santiago 22,
Chile\\
}
\begin{document}

\date{Accepted 2010 August 13. Received 2010 August 12; in original
  form 2010 June 21}

\pagerange{\pageref{firstpage}--\pageref{lastpage}} \pubyear{2002}

\maketitle

\label{firstpage}

\begin{abstract}
\iphas\ is a new high-excitation planetary nebula with remarkable
characteristics.
It consists of a knotty ring expanding at a speed of 28~\kms, and a
fast collimated outflow in the form of faint lobes and caps along the
direction perpendicular to the ring. The expansion speed of the polar caps
is $\sim$100~\kms, and their kinematical age is twice as large as the age
of the ring. 

Time-resolved photometry of the central star of \iphas\ reveals a
sinusoidal modulation with a period of 1.16~days. This is interpreted
as evidence for binarity of the central star, the brightness
variations being related to the orbital motion of an irradiated
companion. This is supported by the spectrum of the central star in
the visible range, which appears to be dominated by emission from the
irradiated zone, consisting of a warm (6000-7000~K) continuum, narrow
C~III, C~IV, and N~III emission lines, and broader lines from a flat
H~I Balmer sequence in emission.

\iphas\ helps to clarify the role of (close) binaries in the formation
and shaping of planetary nebulae. The output of the common-envelope
evolution of the system is a strongly flattened circumstellar mass
deposition, a feature that seems to be distinctive of this kind of
binary system. Also, \iphas\ is among the first post-CE PNe for which
the existence of a high-velocity polar outflow has been
demonstrated. 
Its kinematical age might indicate that the polar outflow is formed
before the common-envelope phase.
This points to mass transfer onto the secondary as the origin, but
alternative explanations are also considered.
\end{abstract}


\begin{figure*}
\includegraphics[width=176mm]{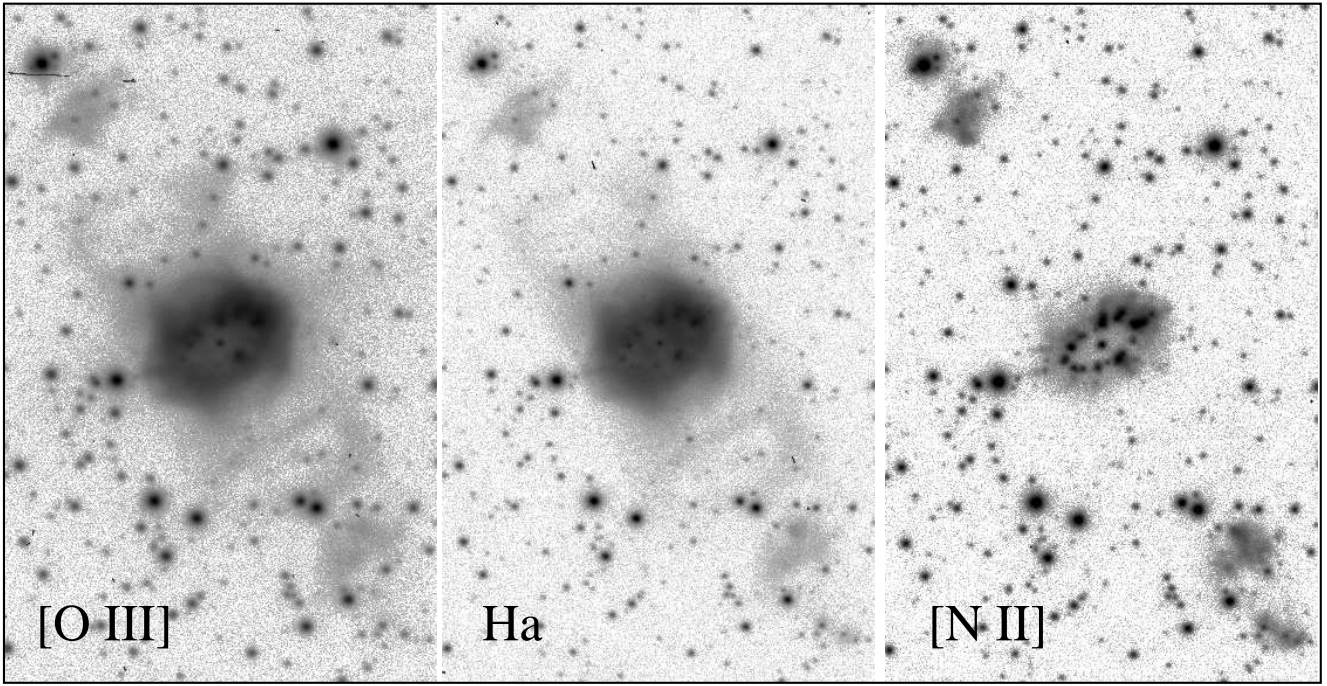}
  \caption{The NOT images of \iphas\, in a log intensity scale. 
The field of view is 70$''$$\times$110$''$ in each  frame. 
North is up and East is left.}
\label{F-image}
\end{figure*}

\section{Introduction}

The IPHAS H$\alpha$ survey of the Northern Galactic plane \citep{d05}
provides a wealth of new information about emission-line
sources. Among them, a large number of planetary nebulae (PNe) are
being discovered \citep{kert09,s10}. Two main objectives can be
generally pursued using the IPHAS data on PNe. The first is to
determine the global properties of PNe, such as their population size in
the Galactic Plane and related issues (stellar death rate, global
yields, relative contribution of the different morphological classes
and their distribution in the disc, etc.).  The second objective is to find 
and study new objects belonging to rare morphological or chemical sub-classes, 
or that have other outstanding
characteristics.  \iphas\ is one of these distinctive new PNe. It is
among the brightest new IPHAS PNe in the list of \citet{s10}, and its
characteristic morphology made it a prime candidate in our ongoing
programme looking for binary central stars based on the emerging
morphological trends identified with nebulae around close binaries
\citep{mis09b}. In the following, we present imaging, spectroscopy,
and time-resolved photometry which allow us to determine 
geometrical, kinematical, physical and chemical properties of the
nebula, and to reveal the existence of a close binary central
star. This provides a new clarifying example of the relevance of
interactions in close binaries for the shaping of PNe -- an important
and long standing problem in the field \citep{dm09}.

\section{Observations}
\label{S-obs}

\begin{figure}
\centering
\includegraphics[width=60mm]{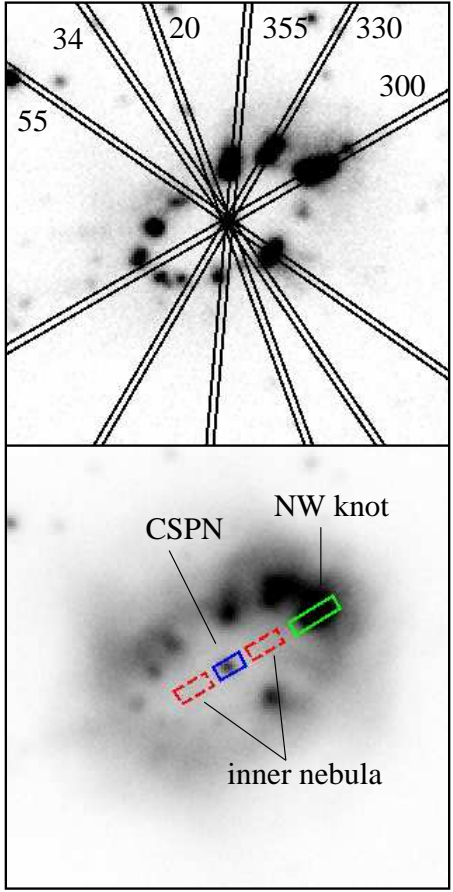}
  \caption{Position of the slits used for spectroscopy at the NOT
    (top) and at the WHT (bottom) are shown over a
    30$''$$\times$30$''$ zoom of the [N~II] and [O~III] images,
    respectively, with linear intensity scale. In the latter, the
    central star (CSPN) and the two regions of the nebula discussed in the
    text are indicated.}
\label{F-imageslit}
\end{figure}

\iphas\ was discovered as a new candidate PN from images of the IPHAS
survey obtained on 2005 (Sabin et al. 2010). Deeper and higher
resolution images were secured at the 2.6m Nordic Optical Telescope
(NOT) and the ALFOSC spectro-imager on la Palma on 3 September 2007.
The filter central wavelengths and full width at half maximum (FWHM),
and the corresponding selected nebular emission lines, were:
6589/9~\AA\ ([N~II] 6583), 6568/8~\AA\ ($H\alpha$ 6563) and
5007/30~\AA\ ([O~III] 5007). Note that the red filters are narrow
enough that no contamination of the [N~II] lines occurs in the
$H\alpha$ filter and vice versa. In each filter, the exposure time was
20 min. The spatial scale of the ALFOSC instrument is
$0''.19$~pix$^{-1}$, and seeing was $0''.6$ FWHM for the $H\alpha$ and
[N~II] images, and $0''.8$ for [O~III].  The NOT images are shown in
Fig.~\ref{F-image}.

With the same instrumentation, medium-resolution long-slit spectra
centred on $H\alpha$ were obtained on the same night and in the
following one. Grism~\#17 was used, which provides a spectral
dispersion of 0.25~\AA~pix$^{-1}$, a resolution of 0.7~\AA\ (FWHM)
with the adopted slit width of $0''.5$, and a coverage from
6375~\AA\ to 6745~\AA.  The slit was positioned through the central
star of the nebula at several position angles, namely at P.A.=20\deg,
34\deg, 55\deg, 300\deg, 330\deg, and 355\deg, in order to map the
knotty inner ring-like structure (Fig.~\ref{F-imageslit}, upper
panel). Exposure times were 10 min, except for the spectrum through
P.A.=+34\deg, where it was 1 hour.

A lower resolution spectrum of the nebula and its central star was
obtained on the 17 July 2007 with the 4.2m~WHT telescope and the
double-arm ISIS spectrograph. The long slit of ISIS was opened to 1
arcsec width and positioned at P.A.=121\deg, roughly along the major
axis of the ring, passing through the central star and the brightest
knot on its NW side (Fig.~\ref{F-imageslit}, lower panel). In the blue
arm of ISIS, grating R300B was used, providing a dispersion of
0.86~\AA~pix$^{-1}$, a resolution of 3~\AA, and a spectral coverage
from 3200 to 5400~\AA. In the red arm, grating R158R gave a dispersion
of 1.82~\AA~pix$^{-1}$, a resolution of 6~\AA, and a spectral coverage
from 5400 to 10000~\AA. Note however that due to the dichroic used to
split the blue and red light, and to vignetting inside the
spectrograph, flux calibration is uncertain in the range from 5300 to
5600~\AA, and above 9200~\AA. Total exposure times were 40 and 60
minutes in the blue and red arms, respectively. Seeing was 0$''$.8.
Several spectrophotometric standards from \citet{o90} were observed
during the night, while arcs were only obtained during daytime, a fact
that limits the precision of the wavelength calibration.

Time-resolved photometry in the SDSS $i$ band of \iphas\ was performed
during 2009 with the 1.2m Mercator telescope and its MEROPE camera,
the 2.5m Isaac Newton Telescope (INT) and its Wide Field Camera (WFC),
and the 0.8m IAC80 and its CAMELOT camera. The time coverage of each
set of images was different, while the individual exposure times were
chosen to ensure a S/N$\sim$120 on the integrated emission from the
central star. In the case of the MERCATOR/MEROPE observations, the
time coverage was split into two blocks separated by several hours to
better sample variability. The log of the photometric observations can
be found in Tab.~\ref{T-photoobs}.  Magnitudes obtained with different
telescope/filter combinations were matched to the INT SDSS $i$
photometric system using field stars. The data are listed
in Tab.~\ref{T-photdata}.

\begin{table}
\centering
\caption{Log of the photometric observations of \iphas.} 
\begin{tabular}[t]{lllcc}
\hline
\,Date & Instrum.  &  Filter &   Exp. time & Coverage   \\    
(2009) &   &           &     [sec]               & [min]    \\    
\hline
Aug 25 &  MEROPE   &   Cousins I   &   200     & 2$\times$25 \\
Aug 28 &  MEROPE   &   Cousins I   &   200     & 2$\times$20 \\
Aug 31 &  MEROPE   &   Cousins I   &   200     & 2$\times$60 \\
Sep 1  &  MEROPE   &   Cousins I   &   200     & 60+15 \\
Sep 2  &  MEROPE   &   Cousins I   &   200     & 75 \\
Sep 10 &  CAMELOT  &   Johnson I   &   600      & 90 \\
Oct 25 &  WFC      &   SDSS    i   &    60      & 180 \\
\hline
\end{tabular}
\label{T-photoobs} 
\end{table} 

All data were reduced using standard routines in {\sc iraf}.

\begin{table}
\centering
\caption{SDSS $i$ magnitudes of the central star of \iphas.  A small
  sample is provided below, the full table being available in the
  electronic version of the article.  }
\begin{tabular}[t]{cll}
\hline
HJD & $i$-mag & err  \\    
\hline
2455069.452360  & 17.90 &  0.02\\ 
2455069.455140  & 17.91 &  0.02\\ 
2455069.457900  & 17.89 &  0.02\\ 
2455069.460670  & 17.87 &  0.02\\ 
2455069.463420  & 17.87 &  0.02\\ 
2455069.466190  & 17.86 &  0.02\\ 
2455069.468930  & 17.89 &  0.02\\ 
2455069.471730  & 17.87 &  0.02\\ 
2455069.535730  & 17.69 &  0.02\\ 
2455069.538470  & 17.69 &  0.02\\ 
2455069.541210  & 17.68 &  0.02\\ 
2455069.543990  & 17.70 &  0.02\\ 
2455069.546760  & 17.71 &  0.02\\ 
 ....           &       &      \\     
\hline
\end{tabular}
\label{T-photdata} 
\end{table} 

\begin{figure}
\centering
\includegraphics[width=65mm]{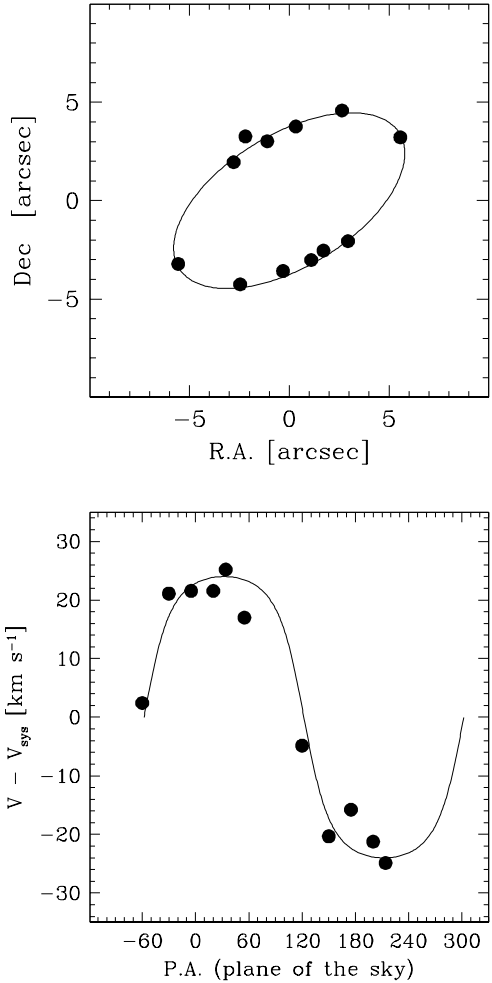}
  \caption{The ring's kinematical fit. Top: dots
    indicate the positions at which radial velocities are measured, i.e.  
    the intersection of the NOT slits with the brightest part of the
    ring. Bottom: dots are the corresponding [N~II] line-of-sight velocities.  
Solid lines show the fit to the data with an inclined, circular expanding ring
    with the parameters quoted in the text.}
\label{F-ringvel}
\end{figure}

\begin{figure}
\centering
\includegraphics[width=80mm]{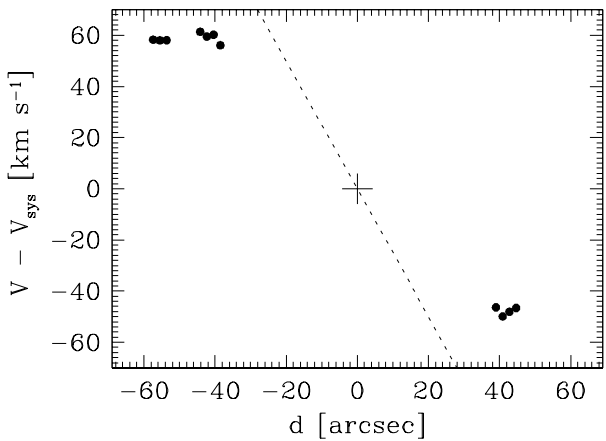}
  \caption{The observed [NII] velocities in the polar caps (dots). The
    dotted line shows the variation of velocities with distance from
    the central star, assuming a proportionality law with a slope
    fixed by the expansion speed of the ring.}
\label{F-jets}
\end{figure}

\section{Morphology, kinematics, geometry and orientation}
\label{S-kin}

The morphology of the nebula (Fig.~\ref{F-image}) is similar in the
H$\alpha$ and [O~III] emission: it consists of a diffuse, elliptical
or perhaps spindle-like inner body, whose brightest regions are in the
form of a knotty elliptical ring with a long axis of about 13 arcsec.
Along the orthogonal direction, faint emission traces a
roughly-cylindrical structure ending in two slightly brighter,
irregular ``polar'' caps at about 1 arcmin from the central star. As
typical of this kind of structure \citep{g01}, the ring's knots and
their faint outward tails, as well as the polar caps are most evident
in the low-ionization [NII] emission. This striking [NII] morphology
led to the object's nickname, ``the Necklace Nebula'' \citep{s10}.

The kinematics and 3D structure of the nebula were studied by means of
the Doppler shift of the [N~II] 6583 line (which has a smaller thermal
broadening than H$\alpha$) in the NOT spectra.
The [N~II] line-of-sight velocities at the position of the ring's
knots
are well fit (Fig.~\ref{F-ringvel}) by a circular expanding ring with
the following parameters: ring radius $r$=6$''$.5$\pm$0$''$.5, inclination to the
plane of the sky $i$=59\deg$\pm$3\deg\ (where $i$=0\deg\ is the ring in the
plane of the sky and $i$=90\deg\ is in the line of sight), major axis
of the ring projected on the sky at P.A.=122\deg$\pm$3\deg, expansion velocity
$V_{ring}$=28$\pm$3~\kms, and systemic velocity $V_{sys}$=45$\pm$2~\kms,
corrected to the Local Standard of Rest. The distance-dependent age of
the ring, assuming no acceleration, is \td$_{ring}$=1100~yr~kpc$^{-1}$.

The [N~II] velocities of the polar caps are shown in
Fig.~\ref{F-jets}.  Assuming that they are located along the polar
axis of the ring, their deprojected mean velocities are
115~\kms\ for the southern cap, and 95~\kms\ for the northern
one. 

Their age/distance parameter \td$_{caps}$ is between 1900 (part of
the southern cap closer to the star) and 2800~yr~kpc$^{-1}$ (outermost
region of the southern cap). This is roughly twice as large as for the
ring. Thus, in the simple hypothesis of ballistic motions the caps
must have been ejected before the ring. The lack of a clear increase
of speed with distance from the central star {\it along} each cap
might also suggest a continuous ejection process that lasted for a
significant length of time, during a now extinct jet phase perhaps.


The limited resolution and depth of the NOT spectra do not allow us to
perform a kinematical study of the faint emission extending from the ring
to the caps.

We conclude that the morphology of the \iphas\ nebula is that of an
``equatorial'' ring, where most of the emission (ionized mass) is
concentrated, from which faint lobes depart, ending in low-ionization 
``polar'' caps.

\begin{table*}
 \centering
  \caption{Observed (F$_{obs}$) and dereddened (F$_{der}$) nebular emission line fluxes,
    relative to H$\beta$=100, and their errors. 
    H Balmer lines include the contribution of
    HeII Pickering lines (see text).}
\begin{tabular}{lcrrrrrrrr}
\hline
  Identification                  & $\lambda_{obs}$    & \multicolumn{4}{c}{Inner nebula}        & \multicolumn{4}{c}{NW knot}             \\
                                  & \AA    & F$_{obs}$ & err\%  & F$_{der}$& err\%     &F$_{obs}$ & err\%    & F$_{der}$ & err\%  \\
\hline                             
HeII 3203.1                         & 3201.1 &     25.6 &        7&     46.8&        9 &    18.1 &        8 &     33.0 &       10 \\[-0.5pt] 
\protect{[}Ne V\protect{]} 3345.4   & 3344.3 &     59.9 &        4&    102.0&        7 &     7.7 &       10 &     13.1 &       12 \\[-0.5pt] 
\protect{[}Ne V\protect{]} 3425.5   & 3424.6 &    173.8 &        3&    286.4&        6 &    19.6 &        4 &     32.2 &        7 \\[-0.5pt] 
O III 3444.1?                       & 3442.5 &          &         &         &          &     3.0 &       15 &      5.0 &       16 \\[-0.5pt] 
\protect{[}O II\protect{]} 3726+3729& 3726.8 &      6.0 &       20&      8.9&       21 &    84.4 &        3 &    125.0 &        5 \\[-0.5pt] 
HI 3750.1                           & 3749.6 &          &         &         &          &     1.8 &       24 &      2.6 &       24 \\[-0.5pt] 
HeI 3756.1blend                     & 3755.8 &          &         &         &          &     2.9 &       18 &      4.3 &       19 \\[-0.5pt] 
HI 3770.6                           & 3770.6 &          &         &         &          &     2.8 &       11 &      4.1 &       12 \\[-0.5pt] 
HI 3797.9                           & 3797.3 &      3.7 &       14&      5.3&       15 &     3.5 &        9 &      5.1 &       10 \\[-0.5pt] 
HI 3835.3                           & 3835.0 &      4.9 &        9&      7.0&       10 &     4.4 &        7 &      6.2 &        8 \\[-0.5pt] 
\protect{[}Ne III\protect{]} 3869.0 & 3868.5 &     25.0 &        4&     35.4&        5 &    78.3 &        3 &    110.7 &        5 \\[-0.5pt] 
HI 3889.0                           & 3888.8 &      7.8 &        7&     10.9&        8 &    10.3 &        5 &     14.4 &        6 \\[-0.5pt] 
\protect{[}Ne III\protect{]}+HeII   & 3968.0 &     18.8 &        5&     25.7&        6 &    36.0 &        3 &     49.1 &        5 \\[-0.5pt] 
HeII 4025.6                         & 4025.8 &          &         &         &          &     1.3 &       24 &      1.7 &       24 \\[-0.5pt] 
\protect{[}S II\protect{]} 4068.6   & 4068.7 &          &         &         &          &     4.0 &        8 &      5.3 &        8 \\[-0.5pt] 
\protect{[}SII\protect{]} 4076.3    & 4076.6 &          &         &         &          &     1.0 &       20 &      1.3 &       20 \\[-0.5pt] 
HI 4101.7                           & 4101.7 &     20.1 &        4&     26.2&        5 &    19.9 &        4 &     25.9 &        5 \\[-0.5pt] 
HeII 4199.8                         & 4200.5 &      1.9 &       25&      2.4&       25 &     1.1 &       25 &      1.4 &       25 \\[-0.5pt] 
HI 4340.5                           & 4340.5 &     39.7 &        3&     47.5&        4 &    38.5 &        3 &     46.0 &        4 \\[-0.5pt] 
\protect{[}O III\protect{]} 4363.2  & 4363.3 &      7.9 &        7&      9.4&        7 &    13.9 &        4 &     16.5 &        4 \\[-0.5pt] 
HeI 4471                            & 4471.8 &          &         &         &          &     2.3 &       11 &      2.6 &       11 \\[-0.5pt] 
HeII 4541.6                         & 4541.5 &      3.6 &       12&      4.0&       12 &     2.6 &       10 &      2.9 &       10 \\[-0.5pt] 
HeII 4685.7                         & 4685.7 &    105.2 &        3&    111.6&        3 &    79.6 &        3 &     84.4 &        3 \\[-0.5pt] 
\protect{[}Ar IV\protect{]} 4711.4  & 4711.6 &     17.1 &        4&     18.0&        4 &     9.3 &        4 &      9.8 &        4 \\[-0.5pt] 
\protect{[}Ar IV\protect{]} 4740.2  & 4740.1 &     12.6 &        5&     13.1&        5 &     6.9 &        5 &      7.1 &        5 \\[-0.5pt] 
HI 4861.4                           & 4861.2 &    100.0 &        3&    100.0&        3 &   100.0 &        3 &    100.0 &        3 \\[-0.5pt] 
\protect{[}O III\protect{]} 4958.9  & 4958.8 &    179.3 &        3&    173.7&        3 &   436.8 &        3 &    423.1 &        3 \\[-0.5pt] 
 \protect{[}O III\protect{]} 5006.8 & 5006.7 &    534.6 &        3&    510.0&        3 &  1280.7 &        3 &   1221.6 &        3 \\[-0.5pt] 
\protect{[}N I\protect{]} 5199      & 5198.6 &          &         &         &          &     3.7 &        9 &      3.4 &        9 \\[-0.5pt] 
HeII 5411.5                         & 5411.9 &      9.4 &       30&      7.9&       30 &     7.5 &       30 &      6.3 &       30 \\[-0.5pt] 
\protect{[}Cl III\protect{]} 5517.7 & 5518.0 &      1.2 &       33&      1.0&       33 &     2.2 &       15 &      1.8 &       15 \\[-0.5pt] 
\protect{[}Cl III\protect{]} 5537.9 & 5538.5 &      0.7 &       47&      0.6&       47 &     1.8 &       19 &      1.5 &       19 \\[-0.5pt] 
\protect{[}N II\protect{]} 5754.64  & 5754.8 &          &         &         &          &     4.2 &        6 &      3.2 &        6 \\[-0.5pt] 
HeI 5875.6                          & 5875.7 &      1.8 &       29&      1.3&       29 &     8.2 &        6 &      6.2 &        6 \\[-0.5pt] 
\protect{[}O I\protect{]} 6300.3    & 6300.1 &          &         &         &          &    19.8 &        3 &     13.4 &        5 \\[-0.5pt] 
\protect{[}S III\protect{]} 6312.1  & 6312.1 &      4.8 &        7&      3.3&        8 &     9.8 &        4 &      6.6 &        6 \\[-0.5pt] 
\protect{[}O I\protect{]} 6363.8    & 6363.7 &          &         &         &          &     6.7 &        4 &      4.4 &        6 \\[-0.5pt] 
\protect{[}Ar V\protect{]} 6434.7   & 6435.2 &      7.2 &        5&      4.7&        7 &     1.1 &       18 &      0.7 &       18 \\[-0.5pt] 
HeII 6527.1                         & 6526.6 &      1.9 &       20&      1.2&       20 &         &          &          &          \\[-0.5pt] 
\protect{[}N II\protect{]} 6548.1   & 6547.9 &      5.3 &        8&      3.4&        9 &    93.7 &        3 &     60.2 &        6 \\[-0.5pt] 
HI 6562.8                           & 6562.7 &    416.8 &        3&    267.1&        6 &   460.0 &        3 &    294.8 &        6 \\[-0.5pt] 
\protect{[}N II\protect{]} 6583.4   & 6583.2 &     12.7 &        4&      8.1&        7 &   298.1 &        3 &    190.2 &        6 \\[-0.5pt] 
HeI+HeII                            & 6679.3 &          &         &         &          &     3.8 &        7 &      2.4 &        8 \\[-0.5pt] 
HeII 6683.2                         & 6682.3 &      2.1 &       18&      1.3&       18 &         &          &          &          \\[-0.5pt] 
\protect{[}S II\protect{]} 6716.4   & 6716.3 &      3.9 &        8&      2.4&       10 &    48.2 &        3 &     29.9 &        6 \\[-0.5pt] 
\protect{[}S II\protect{]} 6730.8   & 6730.6 &      3.3 &       11&      2.1&       12 &    51.7 &        3 &     32.0 &        6 \\[-0.5pt] 
HeII 6890.9                         & 6890.9 &      1.5 &       25&      0.9&       25 &     1.4 &       17 &      0.9 &       18 \\[-0.5pt] 
\protect{[}Ar V\protect{]} 7005.4   & 7005.9 &     17.1 &        4&     10.1&        7 &     2.6 &       10 &      1.5 &       11 \\[-0.5pt] 
HeI 7065                            & 7064.9 &          &         &         &          &     2.6 &        9 &      1.5 &       11 \\[-0.5pt] 
\protect{[}Ar III\protect{]} 7135.8 & 7135.7 &     23.9 &        3&     13.8&        7 &    57.3 &        3 &     33.0 &        7 \\[-0.5pt] 
\protect{[}Ar IV\protect{]}+HeII    & 7175.6 &          &         &         &          &     2.1 &       13 &      1.2 &       14 \\[-0.5pt] 
\protect{[}O II\protect{]} 7319     & 7319.1 &          &         &         &          &     7.1 &        5 &      4.0 &        8 \\[-0.5pt] 
\protect{[}O II\protect{]} 7330     & 7329.6 &          &         &         &          &     5.7 &        5 &      3.2 &        8 \\[-0.5pt] 
\protect{[}Cl IV\protect{]} 7530    & 7529.8 &      1.6 &       24&      0.9&       25 &     1.7 &       18 &      0.9 &       19 \\[-0.5pt] 
HeII 7592.7                         & 7592.3 &      3.1 &       24&      1.7&       25 &     1.9 &       36 &      1.0 &       37 \\[-0.5pt] 
\protect{[}Ar III\protect{]} 7751.1 & 7750.9 &      5.6 &        7&      2.9&       10 &    15.0 &        4 &      7.9 &        8 \\[-0.5pt] 
\protect{[}Cl IV\protect{]} 8046.3  & 8045.9 &      4.9 &        8&      2.5&       11 &     3.2 &        9 &      1.6 &       12 \\[-0.5pt] 
HeII 8236.8                         & 8236.9 &      5.3 &        7&      2.6&       10 &     4.5 &        7 &      2.2 &       11 \\[-0.5pt] 
HI 8545.4                           & 8545.2 &          &         &         &          &     1.1 &       28 &      0.5 &       29 \\[-0.5pt] 
HI 8598.4                           & 8598.5 &      1.6 &       28&      0.8&       29 &     2.1 &       12 &      1.0 &       15 \\[-0.5pt] 
HI 8665.0                           & 8664.6 &      1.2 &       21&      0.6&       23 &     2.4 &       10 &      1.1 &       13 \\[-0.5pt] 
HI 8750.5                           & 8750.2 &      2.3 &       22&      1.1&       24 &     2.4 &       13 &      1.1 &       15 \\[-0.5pt] 
HI 8862.8                           & 8862.1 &          &         &         &          &     3.4 &       13 &      1.6 &       15 \\[-0.5pt] 
HI 9014.9                           & 9014.3 &      3.6 &       15&      1.6&       17 &     4.0 &        9 &      1.8 &       12 \\[-0.5pt] 
\protect{[}S III\protect{]} 9068.6  & 9068.8 &     46.3 &        3&     20.9&        9 &   109.4 &        3 &     49.4 &        9 \\[-0.5pt] 
\hline
\end{tabular}
\label{T-spectra}
\end{table*}

\section{Chemical abundances}
\label{S-chem}

The WHT spectrum was used to compute the physical and chemical
properties of the nebula.  Three regions were considered (see
Fig.\ref{F-imageslit}, bottom panel): the central star (CSPN),
discussed in the next section; the {\em inner nebula} contained within
a distance of 1.6 arcsec to 4.0 arcsec on both sides of the CSPN; and
the bright {\em NW knot} extending from 5.0 to 8.4 arcsec on the NW
side of the CSPN. The emission-line fluxes for the nebular regions are
listed in Tab.~\ref{T-spectra}: the errors quoted include both the
uncertainty in the flux of each line (Poissonian, detector and
background noise) and the determination of the instrumental
sensitivity function for flux calibration.

The logarithmic extinction at $H\beta$, $c_\beta$, was computed from
the hydrogen Balmer decrement by averaging the results from the
observed $H\alpha/H\beta$, $H\gamma/H\beta$, and $H\delta/H\beta$
ratios (weighted by their errors) in the two nebular regions.  Owing
to the high excitation of the nebula, the hydrogen Balmer lines are
contaminated by unresolved HeII Pickering lines at a level of 6\%\ for
the inner nebula, and 4\%\ for the knot. This contribution does not
significantly affect the calculation of $c_\beta$, but must be
subtracted when the elemental abundances are computed (see
below)\footnote{Note that fluxes in Tab.~\ref{T-spectra} are the
  observed ones, and include the contribution of the HeII lines to the
  HI emission.}.  The extinction law by \citet{f04} with R=3.1 was
adopted. The theoretical Balmer line ratios were derived from
\citet{b71} for electron temperatures
as determined below. We obtained $c_\beta=0.55\pm0.06$, corresponding
to $A_V=1.19\pm0.12$.  Line fluxes dereddened using this $c_\beta$
and the Fitzpatrick's extinction law are also listed in
Tab.~\ref{T-spectra}.

The nebula of \iphas\ is of high excitation.  In the inner nebula,
after dereddening [O~III]~5007 is the strongest line followed by
[Ne~V]~3426. HeII~4686 is stronger than H$\beta$, indicating the
highest excitation class (10) according to the scheme set out by \citet{dm90}.
In the NW knot, excitation is slightly lower as HeII and [Ne~V] are
relatively weaker, albeit the [O~III]/H$\beta$ ratio is larger than in
the inner nebula.  This suggests, as confirmed by the chemical
analysis below, that most of the oxygen in the inner nebula is in higher
ionization stages which are not observed in the optical. This also
holds for other elements including nitrogen and sulphur.  In the knot,
neutral species like [O~I] and [N~I] are also present. The mixture of
high and low ionization atoms is only partly caused by the
superposition along the line of sight of the high excitation gas
belonging to the main body of the nebula with the relatively lower
ionization of the ring's knots.

The electron densities \ne\ were computed from the [S~II]~6731/6717
line ratio and, in the inner nebula, also from the [Ar~IV]~4711/4740
ratio. The electron temperature \te\ was computed from the
[O~III]~(5007+4959)/4363 ratio and, in the NW knot, also the
[N~II]~(6583+6548)/5755 ratio. The {\sc nebular} package in IRAF
\citep{sd95} was used, and results are listed in Tab.~\ref{T-chem}. Other
indicators of density ([Cl~III]~5518/5538) and temperature
([O~II]~(3726+3729)/(7320+7330), [S~II]~(6717+6731)/4072,
[S~III]~(9069+9532)/6312) are available in the spectra, but the
associated errors (from both the line measurements and the uncertainty
on temperature/density) are much larger than for the adopted
diagnostic ratios, and were therefore not used.  The [N~II] \te\ in
the knot is 2000~K lower than the [O~III] \te: this is typical of high
excitation PNe \citep{kb94, p98}, as is the quite large value of
\te\ in the inner nebula. Densities in \iphas\ are conversely
relatively low.

\begin{table}
\begin{center}
  \caption{Physical conditions and chemical abundances in the
    \iphas\ nebula.  Total abundances of most metals in the inner nebula are not
    quoted, given their large {\it icf}s (see text).}
\begin{tabular}{lll}
\hline
                                 & Inner nebula  & NW knot     \\
\hline                             
\protect{[}SII\protect{]}  \ne   & 360$^{+380}_{-240}$ cm$^{-3}$ & 820$^{+140}_{-120}$ cm$^{-3}$ \\[5pt]
\protect{[}ArIV\protect{]} \ne   & 370$^{+800}_{-370}$ cm$^{-3}$ &                               \\[5pt]
\protect{[}OIII\protect{]} \te   & 14800$^{+530}_{-460}$ K      & 12960$^{+460}_{-400}$ K       \\[5pt]
\protect{[}NII\protect{]}  \te   & [11000$^{+2000}_{-2000}$ K]$^\ast$                           & 10920$^{+420}_{-380}$ K        \\[10pt]
He$^+$/H$^+$               &     0.011$\pm$0.003   &  0.052$\pm$0.005      \\
He$^{2+}$/H$^+$            &     0.105$\pm$0.016    &  0.077$\pm$0.009     \\
{\bf He/H}                & {\bf 0.116$\pm$0.016}  & {\bf 0.129$\pm$0.010}  \\   
                          & ({\bf 11.06})$^{\natural}$ &({\bf 11.11})$^{\natural}$ \\[3pt]
O$^+$/H$^+$    $\times 10^4$ &     0.025$^{+0.044}_{-0.014}$ &  0.43$\pm$0.13  \\
O$^{2+}$/H$^+$ $\times 10^4$ &     0.59$\pm$0.06           &  2.00$\pm$0.20   \\
{\it icf(O)}               &                             &  1.83$\pm$0.15   \\     
{\bf O/H} $\times 10^4$     &                             &{\bf 4.45$\pm$0.57}\\
                          &                              &({\bf 8.64})$^{\natural}$ \\[3pt]
N$^+$/H$^+$    $\times 10^4$ &    0.015$^{+0.009}_{-0.004}$  &   0.30$\pm$0.03  \\
{\it icf(N)}               &                             &  10.36$\pm$3.40  \\      
{\bf N/H} $\times 10^4$     &                             &{\bf 3.11$\pm$1.07}  \\
                          &                              &({\bf 8.49})$^{\natural}$ \\[3pt]
$\log$(N$^+$/O$^+$)         &{\bf -0.23$\pm$0.60}    &{\bf -0.15$\pm$0.20}  \\[3pt]
Ne$^{2+}$/H$^+$ $\times 10^4$ &   0.098$\pm$0.015     &  0.45$\pm$0.08      \\
Ne$^{4+}$/H$^+$ $\times 10^4$ &    0.71$\pm$0.10      &  0.13$\pm$0.02      \\
{\it icf(Ne)}$^\dag$        &                        &  2.23$\pm$0.36      \\     
{\bf Ne/H} $\times 10^4$$^\dag$  &                    &  {\bf 1.00$\pm$0.24} \\
                          &                         &({\bf 8.00})$^{\natural}$ \\
{\bf Ne/H} $\times 10^4$$^\ddag$ & {\bf 1.21$\pm$0.15} & {\bf 0.87$\pm$0.12} \\
                          &  ({\bf 8.08})$^{\natural}$ &({\bf 7.94})$^{\natural}$ \\[3pt]
Ar$^{2+}$/H$^+$ $\times 10^6$ &   0.57$\pm$0.06       &  1.81$\pm$0.19       \\
Ar$^{3+}$/H$^+$ $\times 10^6$ &   1.10$\pm$0.10       &  0.79$\pm$0.07       \\
Ar$^{4+}$/H$^+$ $\times 10^6$ &   0.94$\pm$0.10       &  0.18$\pm$0.03        \\
{\it icf(Ar)}               &                       &  1.11$\pm$0.04        \\     
{\bf Ar/H} $\times 10^6$     &                      &{\bf 3.08$\pm$0.26}     \\
                          &                        &({\bf 6.49})$^{\natural}$ \\[3pt]
S$^+$/H$^+$    $\times 10^6$ &   0.093$\pm$0.036     &   1.29$\pm$0.21       \\
S$^{2+}$/H$^+$ $\times 10^6$ &   2.33$\pm$0.45       &   7.02$\pm$1.40        \\
{\it icf(S)}               &                       &   1.56$\pm$0.46        \\     
{\bf S/H} $\times 10^6$     &                       &{\bf 13.98$\pm$4.43}    \\
                          &                        &({\bf 7.15})$^{\natural}$ \\[3pt]
Cl$^{2+}$/H$^+$ $\times 10^7$ &   0.37$\pm$0.12       & 1.07$\pm$0.19       \\
Cl$^{3+}$/H$^+$ $\times 10^7$ &   1.01$\pm$0.20       & 0.80$\pm$0.49       \\
{\it icf(Cl)}               &                       & 2.48$\pm$0.31       \\     
{\bf Cl/H} $\times 10^7$     &                       &{\bf 4.64$\pm$ 1.42} \\
                          &                         &({\bf 5.67})$^{\natural}$ \\
\hline
\end{tabular}
\end{center}
$^\ast$ Adopted (see text)\\
$^{\natural}$ Total abundances expressed as $\log (\frac{X}{H}) + 12$ \\
$^\dag$ Ne/H=icf$\times$(Ne$^{2+}$/H$^+$) \citep{kb94}\\
$^\ddag$ Ne/H=1.5$\times$(Ne$^{2+}$/H$^+$+Ne$^{4+}$/H$^+$) \citep{kb94}
\label{T-chem}
\end{table}

Given the temperatures and densities, the ionic abundances relative to
hydrogen for all elements but helium were computed from the line fluxes
relative to H$\beta$ using the {\sc ionic} task in IRAF.  Before the
ionic abundances are computed, the contribution of the HeII 8-4
Pickering line has been subtracted from the H$\beta$ line.  

For the knot's spectrum, we have adopted the observed [OIII] \te\ for
medium and high excitation species (doubly-ionized metals or in higher
ionization stages), and the [NII] \te\ for the lower ionization
species. For the inner nebula, the [NII] \te\ is not determined
directly, and a value of 11000$\pm$2000~K was adopted. This is similar
to the electron temperature in the knot's spectrum, fits the empirical
relationship between \te([NII]) and \te([OIII]) and
the HeII4686 line intensity of equation~2 in \citet{kb94}, and is
slightly larger than predicted by equation~3 of \citet{kb94} and by
\citet{m03}.
The concentration of helium ions relative to hydrogen was derived from the
usual recombination lines assuming case B with the effective
recombination coefficients from \citet{hs87} for the HeII and H$\beta$
lines, and from \citet{b99} for the HeI lines.

Errors were derived by propagating both the errors in the line
measurement and those on the adopted temperature and density. When the
ionic abundance of an element is estimated from several independent
emission lines, the weighted mean and its standard deviation were 
computed.  To obtain the total abundances, we used the ionization
correction factors ({\it icf}s) listed in \citet{ab97}, whose primary
source is \citet{kb94}. For chlorine, we adopt the {\it icf}
formulation by \citet{kh01}.  Errors on the total abundances are
obtained by propagating the errors on the mean ionic abundances as
well as on the {\it icf}s.  Ionic and total abundances, and their
errors, are listed in Tab.~\ref{T-chem}.
Note that the total
abundances in the inner nebula are not quoted for any element except
helium. This is because most of the gas is in high ionization stages which
are not measured, making the {\it icf}s (120 for nitrogen!) and their
errors, and thus the computed abundances, very large.  The problem is
less severe in the lower ionization NW knot.

According to the chemical abundances computed in the ring's knot,
\iphas\ is located just on the side of type I PNe in the He/H vs. N/O
diagram, within the range covered by bipolar PNe \citep{p98}. However,
the N/O and He/H overabundances are mild and their errors are non
negligible. Abundances for the other elements are typical of the
majority of PNe (cf. Tab.~3 in \citealt{p98}, and Tab.~14 in
\citealt{kb94}).

\begin{figure*}
\includegraphics[width=150mm]{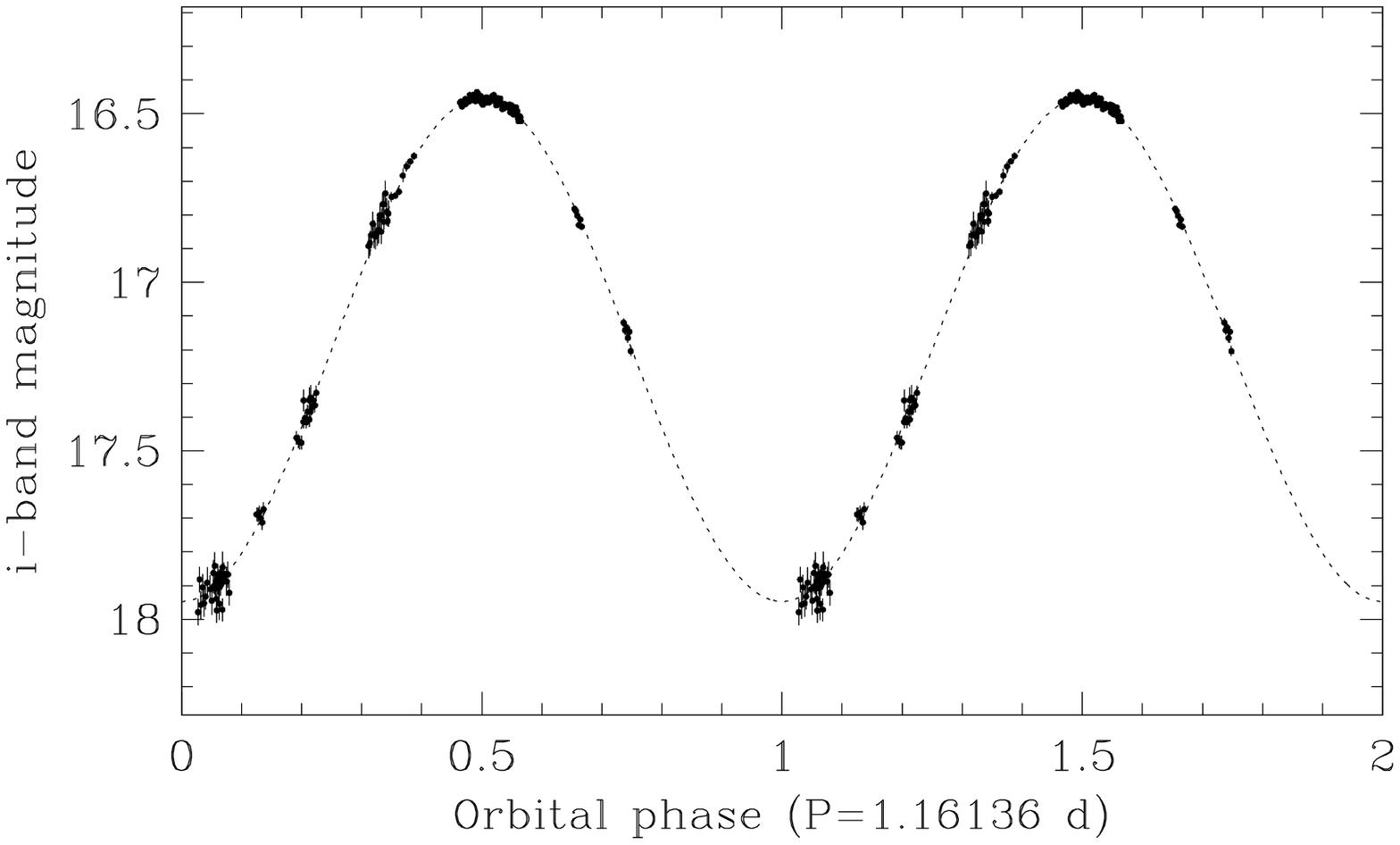}
  \caption{The folded light curve of the \iphas\ central star.}
\label{F-lightcurve}
\end{figure*}

\begin{figure*}
\includegraphics[width=176mm]{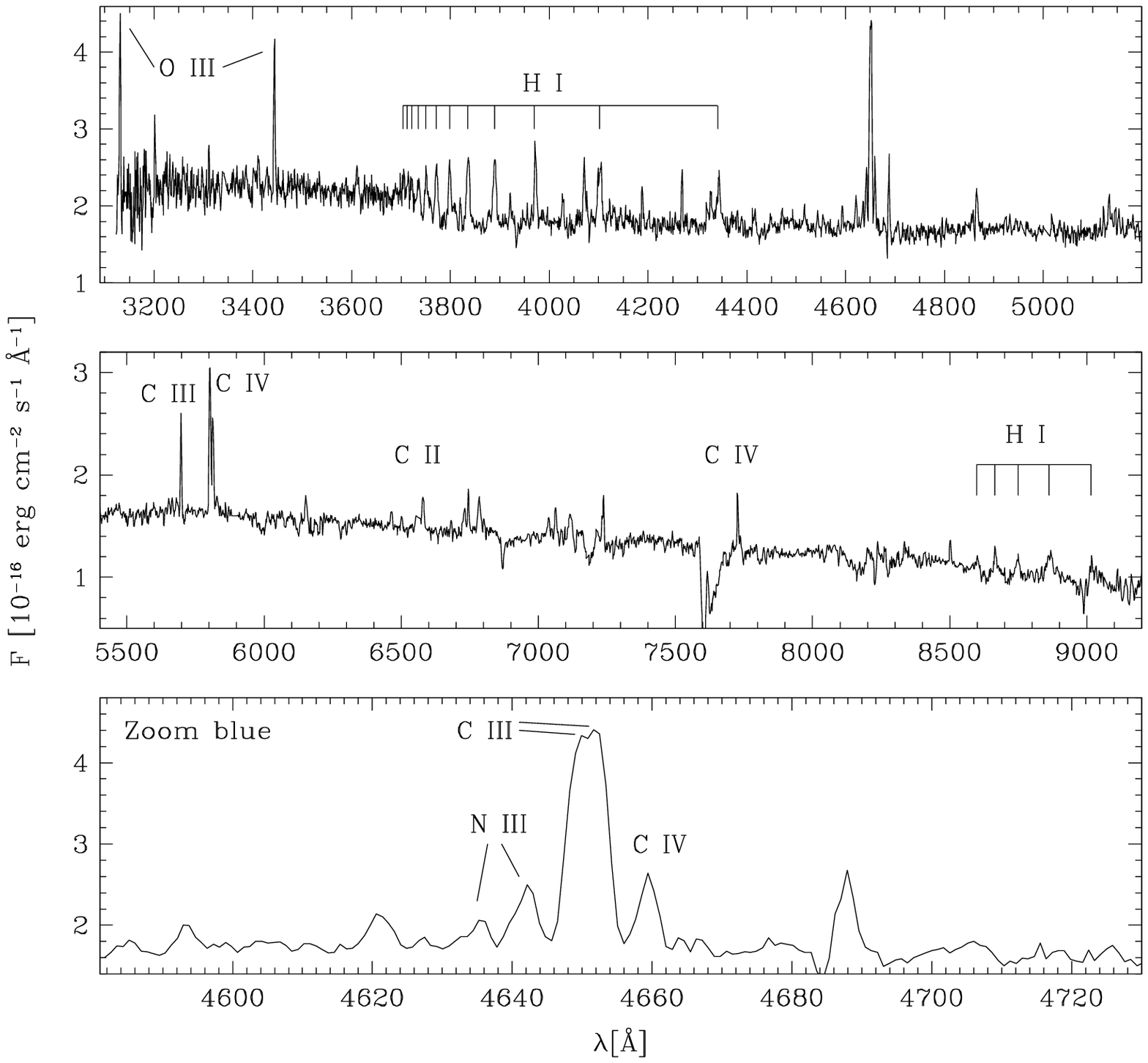}
  \caption{The blue (top) and red (middle) parts of the WHT spectrum
    of the central star of \iphas. At the bottom, a zoom of the blue
    side of the spectrum around the blend at 4650\AA\ is displayed.}
\label{F-csspectrum}
\end{figure*}

\section{The binary central star}

\subsection{Light curve}

The central star's coordinates are R.A.$=$19 43 59.51 and Dec$=$$+$17
09 00.9 (J2000).  Its light curve in the $i$ band from data in
Tab.~\ref{T-photdata}, folded on the period determined below, is
shown in Fig.~\ref{F-lightcurve}.
  
The photometric data set was subjected to a period analysis using
Schwarzenberg-Czerny's \citep{sc96} analysis-of-variance (AOV) method
implemented in {\sc MIDAS}. The AOV periodogram 
shows the strongest peak at $\sim 0.86$ cycle d$^{-1}$, which
corresponds to a period of $1.16$~days. A sine fit to the whole data
set results in $\mathrm{P} = 1.16136 \pm 0.00002$ d, an $i$-band
amplitude (defined as half the peak-to-peak variation) 
of $0.747 \pm 0.002$ mag (among the largest known in this
class of objects), and an average magnitude of $17.199 \pm 0.002$
mag. The fit is shown as a dotted line in Fig.~\ref{F-lightcurve}.
Uncertainties quoted above are the formal errors of the fit: a better
sampling of the curve at all phases is needed to refine the period and
fix the error.  In any case, we take the observed CSPN variability as
a clear indication of binarity, the brightness modulation being the
result of observing the irradiated side of a companion, heated by the
hot core that ejected the PN, at different angles during the orbital
motion. This is supported by the spectrum presented in next section.
Within this hypothesis, maximum brightness occurs at $2455075.7780
\pm0.0007$ (HJD), at orbital phase $\Phi$$=$$0.5$ i.e. at inferior
conjunction of the hot component.

The amplitude of reflection effect seen here is at the high end, but
compatible with those seen in a range of young post common envelope 
binaries \citep{aung}.

\subsection{The irradiated spectrum}

The spectrum of the CSPN is shown in Fig.~\ref{F-csspectrum}.  It
consists of a continuum slowly declining with increasing wavelength, and a 
rich set of emission-lines.  In the figure, the contribution from the extended
nebula was subtracted by interpolating nebular regions on the opposite
sides of the CSPN and as close to it as possible.  The subtraction is
generally accurate, except at the wavelengths of the brightest nebular
lines (mainly HeII4686, H$\beta$ and H$\alpha$).  At the resolution
(3~\AA) and S/R ratio ($\sim$25) of the blue side of our spectrum, no
absorption lines from the hot ionizing CSPN are detected (the feature
at HeII~4686 is likely a nebula subtraction residual). The phase of
the orbital period for this spectrum is not known, as the period
parameters are not precise enough to trace it back to 2007.


As in similar close binary CSPNs (e.g. BE UMa, P=2.3 days,
\citealt{fj94}), the high order lines of the HI Balmer and (fainter)
Paschen sequences are observed in emission. The Balmer decrement,
which has a better S/N ratio and resolution, is remarkably flat from
H$\gamma$ to the highest order transitions.  Information for H$\beta$
and H$\alpha$ is limited by the uncertainty in the subtraction of the
overlapping nebula, as mentioned above, but the two lines are
definitely weaker than H$\delta$. In BE UMa, this flat Balmer
decrement is explained as an optical depth effect in the dense
irradiated atmosphere of the secondary \citep{fj94}. The high-order
Balmer lines are resolved (FWHM$\sim$5~\AA\ after correcting for the
instrumental width) which can be explained by Stark broadening.  As
in BE UMa the Balmer jump is seen in emission.

The spectrum shows a number of narrow (unresolved) emission lines,
which are also assumed to come from the irradiated side of the
secondary.  The most prominent group is the blend around
4650~\AA\ (bottom panel of Fig.~\ref{F-csspectrum}). Given the
presence of strong OIII lines at 3131 and 3444~\AA, the origin of some
lines of this blend may be explained by the Bowen fluorescence
mechanism \citep{b34}, which would result in the formation of the
N~III lines observed at 4635 and 4641~\AA. The peak of the blend is
resolved in two components which are identified as the 3S-3Po C~III
transitions at 4649 and 4652~\AA. This is preferred to the alternative
identification of lower excitation O~II lines (but a contribution of
both is not excluded), because of a better wavelength match, and
because the C~III~5696, C~IV~4658, C~IV~5801 and C~IV~5812 lines are also
observed, showing a quite high degree of excitation in the irradiated
spectrum of \iphas. This is higher than in e.g. BE UMa, where no C~IV
lines are detected, but similar to e.g. V477~Lyrae \citep{pb94} and
ETHOS1, another new PN with a close binary nucleus \citep{mis10}.
Other C~III and C~II lines, as well as low and high-excitation narrow
lines of other elements, are tentatively identified in the WHT
spectrum of \iphas. Given the limited spectral resolution, and the
uncertainty in the Doppler shift (and hence of the line
identification) introduced by the unknown contribution of the orbital
motion, we defer a more detailed analysis of the emission line
spectrum when higher quality data and a radial velocity curve are
obtained.

The continuum, dereddened using $A_V=1.19$ (Sect.~\ref{S-chem}),
cannot be fit by a single black body over the whole observed spectral
range (3200 to 9200~\AA). A single black body with T=8000~K provides a
reasonably good fit only for $\lambda$$>$5000~\AA, while a good fit to
the whole spectral range is obtained with a black body temperature
between 6000 and 7000~K (typical of an F-type star) plus an additional
much hotter component -- presumably the ionizing/irradiating primary
-- whose contribution would dominate the emission at wavelengths
shorter than 4000-5000~\AA. The temperature estimated for the
reprocessed component of the secondary atmosphere is similar to that
determined for the companion in BE~UMa.

\section{Distance, mass, size and age}
\label{S-dist}

The distance to \iphas\ can be estimated using the relation between
the H$\alpha$ surface brightness and radius of the nebula
\citep{f06}. This can be seen as a variant of the mass--radius
relationship \citep{d82}, but it has been carefully calibrated over a
large range of sizes and surface brightnesses and for different
classes of PNe. The radius of \iphas, computed as the geometric mean
of the major and minor semi-axes of the inner elliptical body of the
nebula defined at 10\%\ peak brightness, is $r$$=$10$''$.5.  The
dereddened mean surface brightness of the nebula within this ellipse
is $S$(H$\alpha$)~$=$~6.1$\pm$1.3~10$^{-15}$~erg~cm$^{-2}$~s$^{-1}$~arcsec$^{-2}$.
It was computed by combining the flux-calibrated H$\alpha$+[NII] IPHAS 
images, the narrow band NOT images which separate the [NII]6583 and
H$\alpha$ emission, and the flux-calibrated WHT spectra.
These figures, included in the updated log-log correlation between
radius and surface brightness for the subsample of high-excitation,
optically-thin PNe\footnote{Note that this subsample includes a number of PNe 
with close binary CSPNs \citep{f07}.} like \iphas\ 
result in a distance $D$=4.6$\pm$1.1~kpc. 
The quoted error includes the measured uncertainties on $S$(H$\alpha$), 
combined in quadrature with the dispersion in the correlation \citep{f08}.

\iphas\ is located close to the Galactic plane ($b$=$-$3.37\deg) at
longitude $l$=54.16\deg.  The systemic velocity of the nebular ring
($V_{LSR}=$+45~\kms, see Sect.~\ref{S-kin}), assuming that the nebula
participates in the general circular rotation around the Galactic
centre, would imply a lower limit of 1.6~kpc for the distance, at
1$\sigma$ of the estimated error which is mainly due to the velocity
dispersion ellipsoids of stars in the Galaxy \citep{nordstrom}.

For a distance of 4.6~kpc, the ionized mass of its nebula computed
from the total H$\alpha$ flux and for a standard choice of the filling
factor (0.4, with a 100\%\ uncertainty due to the highly aspherical
geometry of the nebula) is 0.06$\pm$0.03~M$_\odot$, within the range
of masses determined for other PNe \citep{f10}. The total size of the
nebula, including the faint polar extensions, is considerable,
2.7~pc. The kinematical age of the ring would be 5000 years, and the
maximum age of the polar caps would be 13000 years.

\section{Discussion}

\iphas\ is a new Galactic PN with remarkable properties. Its
IAU-approved name is PN~G054.2-03.4.  Analysis of the images and the
[N II] line Doppler shifts indicate that its symmetry axis is inclined
to the line of sight by 59\deg, that most of the gas was ejected in an
``equatorial'' direction corresponding to the observed knotty ring,
and that faint ``polar'' extensions culminate in low-ionization caps
travelling at a speed of roughly 100~\kms. The kinematical data would
suggest a sequence of ejection from the progenitor consisting of a
(continuous?) polar outflow during several 10$^3$~yr, followed, a few
10$^3$ years later, by the ejection of the envelope {along the
  equatorial direction}.  A similar age difference between the
high-velocity polar outflow and the inner nebula is computed for Abell
63 \citep{mi07} and Hb~12 \citep{v09} which have a confirmed or
suspected binary CSPN, respectively. This picture is however based on
the simplistic assumption of ballistic motion, and does not take into
account acceleration/deceleration due to the hydrodynamical evolution
of the structures, as well as possible slowing down of gas with time
by interaction with the environment, which would reduce the age
difference with respect to the main equatorial outflow. Note also that
this result apparently contradicts the finding that in proto-PNe jets
are produced slightly later than equatorial torii \citep{hu07}. These
molecular tori expand at a lower speed than the ionized ring of
\iphas, which might indicate different ejection processes or a
significant acceleration of the gas after photoionization and the fast
wind from the central star arise.  However, the two samples do not
seem to be evolutionary linked as no evidence for close binary central
stars are found in the proto-PN objects \citep{hr10}.


The main result of this study is the discovery that the central star
of \iphas is a close binary. Its orbital period is 1.16~days.  This
lies at the longer tail of the binary CSPN period distribution
\citep{mis09a}, and implies that the system went through
common--envelope (CE) evolution before the ejection of the PN, which
made the orbit shrink to the present value. The recent discovery of a
significant number of PN binary central stars \citep{mis09a}, provides
the base for a better understanding of the effects of binarity on the
mass loss processes at the end of stellar evolution. \iphas\ supports
the increasing evidence in PNe (but not in proto-PNe, see above) of a
correlation between the appearance of morphological features like
rings and collimated high-velocity outflows with the presence of a
close binary central star \citep{mis09b}.  Strictly speaking, Abell~63
\citep{mi07} and \iphas\ are the only PNe with a close binary central
star for which the presence of a high-velocity collimated outflow
along the symmetry axis of the main nebula (i.e. the polar directions)
has been demonstrated. In NGC~6337, a supersonic collimated outflow is
also detected, but its orientation with respect to the main nebular
ring is not completely clear \citep{c00,ga09}. Other close binary PNe
with similar polar outflows are likely to be Sab~41 \citep{mis09b} and
ETHOS1 \citep{mis10}, but kinematical data are needed to confirm this.
Most of the other cases quoted in the literature \citep{dm09}, like
the well studied object, K~1--2, are instead dubious identifications
\citep{c99,mis09b}, mainly because of the lack of convincing evidence
for large expansion speeds.

Another notable characteristic of \iphas\ is its prominent equatorial
ring, a morphological feature that appears to be distinctive of PNe
with close binary central stars \citep{mis09b}.  Clear examples are
the already cited NGC~6337 and Sab~41, as well as Hen~2--428
(Santander--Garc\'\i a et al. in preparation). SuWt~2 also falls in
this morphological class, but the nature of the central system (binary
or triple?) is less clear \citep{jo10}. Note that these marked
equatorial outflows are also observed in much wider binaries like the
symbiotic Miras \citep[e.g. Hen 2-147, ][]{sa07}, as well as in massive
stars \citep{sm07} including SN1987A \citep{pl95}.

Models predict that a sudden envelope ejection results from CE
evolution and that it is strongly concentrated toward the orbital
plane \citep{s98}. This would correspond to the ``equatorial''
rings/tori observed in a number of these PNe.  As discussed for
NGC~6337 \citep{c00}, the clumpy appearance of the ring of \iphas, in
the form of low-ionization knots with outward-facing radial tails, can
be explained as density fluctuations created during the envelope
ejection and later exacerbated by the action of the expanding
ionization front or the post--AGB fast wind. Alternatively, the ring's
knots would be created by radiative shocks leading to the
fragmentation of the shell without involving previous density
inhomogeneities. The polar outflow 
could then be the result of inertial confinement of the gas, forced by
the flattened deposition of the envelope, when further shaping of the
PN takes place under the action of a fast wind from the post--AGB
primary \citep{i92}.  Additionally/alternatively, a strong magnetic
field in the AGB envelope, especially if spun up by the gain of
orbital angular momentum during the CE phase, might play a role in the
shaping outflow both in the equatorial plane and in the polar
directions \citep{bl00, sh92}.  In such cases, the polar outflow would
be produced together with the ring (or later), and therefore the
apparent difference in their kinematical ages in \iphas\ (and Abell~63
and Hb~12) would not reflect a real age difference.

Another possibility is that the high-velocity polar outflow is
produced {\em before} the CE phase from an accretion disc around the
secondary, which can accrete matter from the wind of the primary or
via Roche-lobe overflow \citep{sok98,sr00}.  This would explain why
the polar outflow has a kinematical age larger than that of the ring.


In all models above, the inclination of the orbit of the
\iphas\ CSPN would be the same as of the nebular ring, namely 59\deg.

The radiation from the hot CSPN causes strong irradiation effects in
the illuminated side of the secondary. They dominate the spectrum
presented in this paper, and consist of narrow lines of metals with
ionization potential as high as 64.5eV (C~IV), relatively broad HI
emission in a flat Balmer sequence, and a continuum in the visible
range that is fit by a black body with temperature around 6000-7000~K
plus a much hotter component.  It is clear that these remarkable
characteristics invite a more detailed study of the central source,
aimed at detecting the spectral signature of the hot ionizing star,
determining the radial velocity curve for both components, and
modelling the irradiation spectrum to derive the main parameters of
the two stars.  This
is considered the first priority for the future.  

With the data presently available, a robust upper limit of
1.0~M$_\odot$ for the mass of the secondary can be estimated by
converting the apparent magnitude in the $i$ band of the CSPN at
minimum (corrected for the adopted interstellar reddening and adopting
a distance of 4.6~kpc), to an upper limit to the absolute
magnitude of the non-irradiated side of the secondary.
Assuming a mass between 0.5 to 1.0~M$_\odot$ for the hot post-AGB
star, the observed photometric period of 1.16~days implies the
separation of the two stars to be between 4 and 6 solar radii. In no
case the secondary fills its Roche lobe if it is a main-sequence star,
and no mass transfer to the post-AGB star presently occurs in the
system.

\section*{Acknowledgments}

Based on observations obtained with: the 4.2m~WHT and the 2.5m~INT
telescopes of the Isaac Newton Group of Telescopes, the 2.6m Nordic
Optical Telescope operated by NOTSA, the 1.2m Mercator Telescope
operated by the Flemish Community, and the 0.8m~IAC80 telescope,
operating on the islands of La Palma and Tenerife at the Spanish
Observatories of the Roque de Los Muchachos and Teide of the Instituto
de Astrof\'\i sica de Canarias. The WHT data were obtained as part of
the International Time Programme awarded to the IPHAS collaboration.

The work of RLMC, LS, MSG, AMR, and KV has been supported by the
Spanish Ministry of Science and Innovation (MICINN) under the grant
AYA2007-66804.
We are grateful to Orsola De Marco and Noam Soker for fruitful
discussion.

\label{lastpage}

\begin{thebibliography}{99}
\bibitem[\protect\citeauthoryear{Alexander \& Balick}{1997}]{ab97} 
Alexander, J., Balick, B. 1997, AJ, 114, 713
\bibitem[\protect\citeauthoryear{Aungwerojwit et al.}{2007}]{aung}
Aungwerojwit, A., Gänsicke, B. T., Rodríguez-Gil, P., Hagen, H.-J., 
Giannakis, O., Papadimitriou, C., Allende Prieto, C., Engels, D. 2007,	
A\&A, 469, 207
\bibitem[\protect\citeauthoryear{Benjamin et al.}{1999}]{b99} 
Benjamin, R.A., Skillman, E.D., Smits, D.P. 1999, ApJ, 514, 307
\bibitem[\protect\citeauthoryear{Blackman et al.}{2000}]{bl00}
Blackman, E.G., Frank, A., Markiel, J.A., Thomas, J.H., Van Horn, H.M. 
2000, Nature, 409, 485
\bibitem[\protect\citeauthoryear{Blocklehurst}{1971}]{b71} 
Brocklehurst, M., 1971, MNRAS, 153, 471
\bibitem[\protect\citeauthoryear{Bowen}{1934}]{b34} 
Bowen, I.S. 1934, PASP, 46, 146
\bibitem[\protect\citeauthoryear{Condon et al.}{1998}]{con98} 
Condon, J.J., Cotton, W.D., Greisen, E.W., Yin, Q.F., Perley, R.A., Taylor, G.B., Broderick, 
J.J. 1998, AJ, 115, 1693 
\bibitem[\protect\citeauthoryear{Corradi et al.}{1999}]{c99} 
Corradi, R.L.M., Perinotto, M., Villaver, E. Mampaso, A. Gon\c calves, D.R.
1999, ApH, 523, 821
\bibitem[\protect\citeauthoryear{Corradi et al.}{2000}]{c00} 
Corradi, R.L.M., Gon\c calves, D.R., Villaver, E., Mampaso, A., Perinotto, M.,
Schwarz, H.E., Zanin, C. 2000, ApJ, 535, 823
\bibitem[\protect\citeauthoryear{e.g. Daub}{1982}]{d82} 
Daub, C.T., 1982, ApJ, 260, 612 
\bibitem[\protect\citeauthoryear{see De Marco}{2009}]{dm09} 
De Marco, O. 2009, PASP, 121, 316
\bibitem[\protect\citeauthoryear{Dopita \& Meatheringham}{1990}]{dm90} 
Dopita, M.A., Meatheringham, S.J. 1990, ApJ, 357, 140
\bibitem[\protect\citeauthoryear{Drew et al.}{2005}]{d05} 
Drew, J., Greimel, R., Irwin, M.J., et al. 2005, MNRAS, 362, 753 
\bibitem[\protect\citeauthoryear{Ferguson \& James}{1994}]{fj94}
Ferguson, D.H., James, T.A. 1994, ApJS, 94, 723 
\bibitem[\protect\citeauthoryear{Frew}{2008}]{f08}
Frew, D.J., 2008, hD thesis, Macquarie University, Sydney, Australia
\bibitem[\protect\citeauthoryear{Frew \& Parker}{2006}]{f06}
Frew, D.J., Parker, Q.A. 2006, in IAU Symp. 234, Planetary
Nebulae in our Galaxy and Beyond, M.J. Barlow \& R.H. Mendez eds., p. 49
\bibitem[\protect\citeauthoryear{Frew \& Parker}{2007}]{f07}
Frew, D.J., Parker, Q.A.  2007, in
Asymmetrical Planetary Nebulae IV, IAC Electronic Pub., Eds, Corradi,
R.L.M., Manchado, A. \& Soker, N., p. 475
\bibitem[\protect\citeauthoryear{Frew \& Parker}{2010}]{f10}
Frew, D.J., Parker, Q.A.  2010, PASA, 27, 129
\bibitem[\protect\citeauthoryear{Fitzpatrick}{2004}]{f04} 
Fitzpatrick E.L., 2004, in ``Astrophysics of Dust'', 
Witt A.N., Clayton G.C. \& Draine B.T. eds., ASP Conf. Ser., 
Vol. 309, p. 33
\bibitem[\protect\citeauthoryear{Garc\'\i a-D\'\i az et al.}{2009}]{ga09}
Garc\'\i a-D\'\i az, Ma.T., Clark, D.M., L\'opez, J.A., Steffen, W., 
Richer, M.G. 2009, ApJ, 699, 1633
\bibitem[\protect\citeauthoryear{Giammanco et al.}{2010}]{g10}
Giammanco, C., Sale, S.E., Corradi, R.L.M., et al.  2010, A\&A, submitted
\bibitem[\protect\citeauthoryear{Gon\c calves et al.}{2001}]{g01} 
Gon\c calves, D.R., Corradi, R.L.M., Mampaso, A. 2001, ApJ, 547, 302
\bibitem[\protect\citeauthoryear{Huggins}{2007}]{hu07} 
Huggins, P.J. 2007, ApJ, 663, 342
\bibitem[\protect\citeauthoryear{Hrivnak}{2010}]{hr10} 
Hrivnak, B., 2010, Proceedings of the conference 
``Asymmetrical planetary nebulae V. 
The shaping of stellar ejecta'', in preparation
\bibitem[\protect\citeauthoryear{Hummer \& Storey}{1987}]{hs87} 
Hummer, D.G., Storey, P.J. 1987, MNRAS 224, 801
\bibitem[\protect\citeauthoryear{Jones et al.}{2010}]{jo10} 
Jones, D., Lloyd, M., Mitchell, D.L., Pollacco, D.L., O'Brien, T.J., Vaytet, 
N.M.H. 2010, MNRAS, 401, 405
\bibitem[\protect\citeauthoryear{Kingsburg \& Barlow}{1994}]{kb94} 
Kingsburgh, R.L., Barlow, M.J. 1994 MNRAS, 271, 257
\bibitem[\protect\citeauthoryear{Kwitter \& Henry}{2001}]{kh01} 
Kwitter, K.B., Henry, R.B.C. 2001, ApJ, 562, 804
\bibitem[\protect\citeauthoryear{Icke et al.}{1992}]{i92}
Icke, V., Mellema, G., Balick, B., Eulderink, F., Frank, A. 1992,
Nature, 355, 524
\bibitem[\protect\citeauthoryear{e.g. Livio}{2000}]{l00} 
Livio, M. 2000, in Asymmetrical Planetary Nebulae II: From Origins to 
Microstructures, ASP Conf. Ser., Vol. 199, p. 243\
\bibitem[\protect\citeauthoryear{Magrini et al.}{2003}]{m03} 
Magrini, L., Perinotto, M., Corradi, R.L.M., Mampaso, A. 2003, A\&A, 400, 511
\bibitem[\protect\citeauthoryear{Miszalski et al.}{2009a}]{mis09a} 
Miszalski, B., Acker, A., Moffat, A.F.J., Parker, Q.A., Udalski, A. 2009a, 
A\&A, 496, 813
\bibitem[\protect\citeauthoryear{Miszalski et al.}{2009b}]{mis09b} 
Miszalski, B., Acker, A., Parker, Q.A., Moffat, A.F.J. 2009b, A\&A, 505, 249
\bibitem[\protect\citeauthoryear{Miszalski et al.}{2010}]{mis10} 
Miszalski, B. et al. 2010, MNRAS, submitted
\bibitem[\protect\citeauthoryear{Mitchell et al.}{2007}]{mi07} 
Mitchell, D.L., Pollacco, D., O'Brien, T.J., et al.
2007, MNRAS, 374, 1404
\bibitem[\protect\citeauthoryear{Nordstr{\"o}m et al.}{2004}]{nordstrom} 
Nordstr{\"o}m, B., Mayor, M., Andersen, et al. 2004, A\&A, 418, 989
\bibitem[\protect\citeauthoryear{Oke}{1990}]{o90} 
Oke, T.R. 1990, AJ, 99, 1621
\bibitem[\protect\citeauthoryear{Perinotto \& Corradi}{1998}]{p98} 
Perinotto, M., Corradi R.L.M., 1998, A\&A, 332, 721
\bibitem[\protect\citeauthoryear{Plait et al.}{1995}]{pl95} 
Plait, P.C., Lundqvist, P., Chevalier, R.A., Kirshner, R.P. 1995, ApJ, 439, 370
\bibitem[\protect\citeauthoryear{Pollacco \& Bell}{1994}]{pb94} 
Pollacco, D.L., Bell, S.A. 1994, MNRAS, 267, 452
\bibitem[\protect\citeauthoryear{Sabin et al.}{2010}]{s10} 
Sabin L., et al. 2010, MNRAS, in preparation
\bibitem[\protect\citeauthoryear{Santander--Garc\'\i a et al.}{2007}]{sa07} 
Santander--Garc\'\i a, M., Corradi, R.L.M., Whitelock, P.A., Munari, U., 
Mampaso, A., Marang, F., Boffi, F., Livio, M. 2007,  A\&A, 465, 481
\bibitem[\protect\citeauthoryear{Sandquist et al.}{1998}]{s98} 
Sandquist, E.L., Taam, R.E., Chen, X., Bodenheimer, P., Burkert,
A. 1998, ApJ, 500, 909
\bibitem[\protect\citeauthoryear{Schwarzenberg-Czerny}{1996}]{sc96}
Schwarzenberg-Czerny, A. 1996, ApJ, 460, L107
\bibitem[\protect\citeauthoryear{Shaw \& Dufour}{1995}]{sd95} 
Shaw, R.A., Dufour, R.J 1995, PASP, 107, 896
\bibitem[\protect\citeauthoryear{Smith, Bally, \& Walawender}{2007}]{sm07} 
Smith, N., Bally, J., Walawender, J.  2007, ApJ, 134, 846
\bibitem[\protect\citeauthoryear{Soker}{1998}]{sok98} 
Soker, N. 1998, ApJ 496, 833
\bibitem[\protect\citeauthoryear{but see Soker \& Harpaz}{1992}]{sh92} 
Soker, N., Harpaz, A. 1992, PASP, 104, 923
\bibitem[\protect\citeauthoryear{Soker \& Rappaport}{2000}]{sr00} 
Soker, N., Rappaport, S. 2000, ApJ 538, 241
\bibitem[\protect\citeauthoryear{Vaytet et al.}{2009}]{v09}
Vaytet, N.M.H., Rushton, A.P., LLoyd, M., et al. 2009, MNRAS, 398, 385
\bibitem[\protect\citeauthoryear{Viironen et al.}{2009}]{kert09} 
Viironen, K.,  Greimel, R., Corradi, R.L.M. et al. 2009, A\&A, 504, 291
\end{thebibliography}
\end{document}